\documentclass[american,prl,twocolumn]{revtex4-1}
\usepackage[T1]{fontenc}
\usepackage[latin9]{inputenc}
\usepackage{amsmath}
\usepackage{graphicx}

\makeatletter

\providecommand{\tabularnewline}{\\}

\makeatother

\usepackage{babel}
\begin{document}

\title{A nested polyhedra model of isotropic MHD turbulence}

\author{Ö. D. Gürcan$^{1,2}$}

\affiliation{$^{1}$ CNRS, Laboratoire de Physique des Plasmas, Ecole Polytechnique,
Palaiseau}

\affiliation{$^{2}$ Sorbonne Universités, UPMC Univ Paris 06, Paris}
\begin{abstract}
A nested polyhedra model has been developed for magnetohydrodynamic
(MHD) turbulence. Driving only the velocity field at large scales
with random, divergence free forcing results in a clear, stationary
$k^{-5/3}$ spectrum for both kinetic and magnetic energies. Since
the model naturaly effaces disparate scale interactions, does not
have a guide field and avoids injecting any sign of helicity by random
forcing, the resulting three dimensional $k$-spectrum is statistically
isotropic. The strengths and weaknesses of the model are demonstrated
by considering large or small magnetic Prandtl numbers. It was also
observed that the time scale for the equipartition offset with
those of the smallest scales shows a $k^{-1/2}$ scaling.
\end{abstract}
\maketitle
MHD turbulence has been studied in great detail in the past\citep{biskamp:book},
in particular due to its relevance for space applications such as
solar wind turbulence\citep{alexandrova:09}. In the absence of external,
or self generated mean magnetic fields, MHD turbulence tends to be
isotropic\citep{muller:05}. While in nature mean magnetic fields
abound, the statistically isotropic case, is interesting in its own
right, which may be relevant when the background fields are sufficiently
weak.

Nested polyhedra models were introduced recently as self-similar,
spherically symmetric decimations of Fourier space using complete
triangles in Navier-Stokes turbulence\citep{gurcan:17}. In these,
the wave-vector domain is discretized using nested, alternating icosahedron
dodecahedron pairs that are organized in such a way that wavevectors
that are represented by the vertices of these objects always form
complete triads between neighboring scales. They naturally respect
the conservation laws of the original system and since the discretization
is seperated from the formulation of the equations, they are straightforward
to develop for different systems. Here we show a similar model developed
for MHD system of equations. The result is a model that describes
the three dimensional spectral evolution of MHD turbulence , which
in principle has the ability to represent anisotropy. Since there
is no source of anisotropy however, the resulting turbulence is isotropic.
\begin{figure}
\begin{centering}
\includegraphics[width=1\columnwidth]{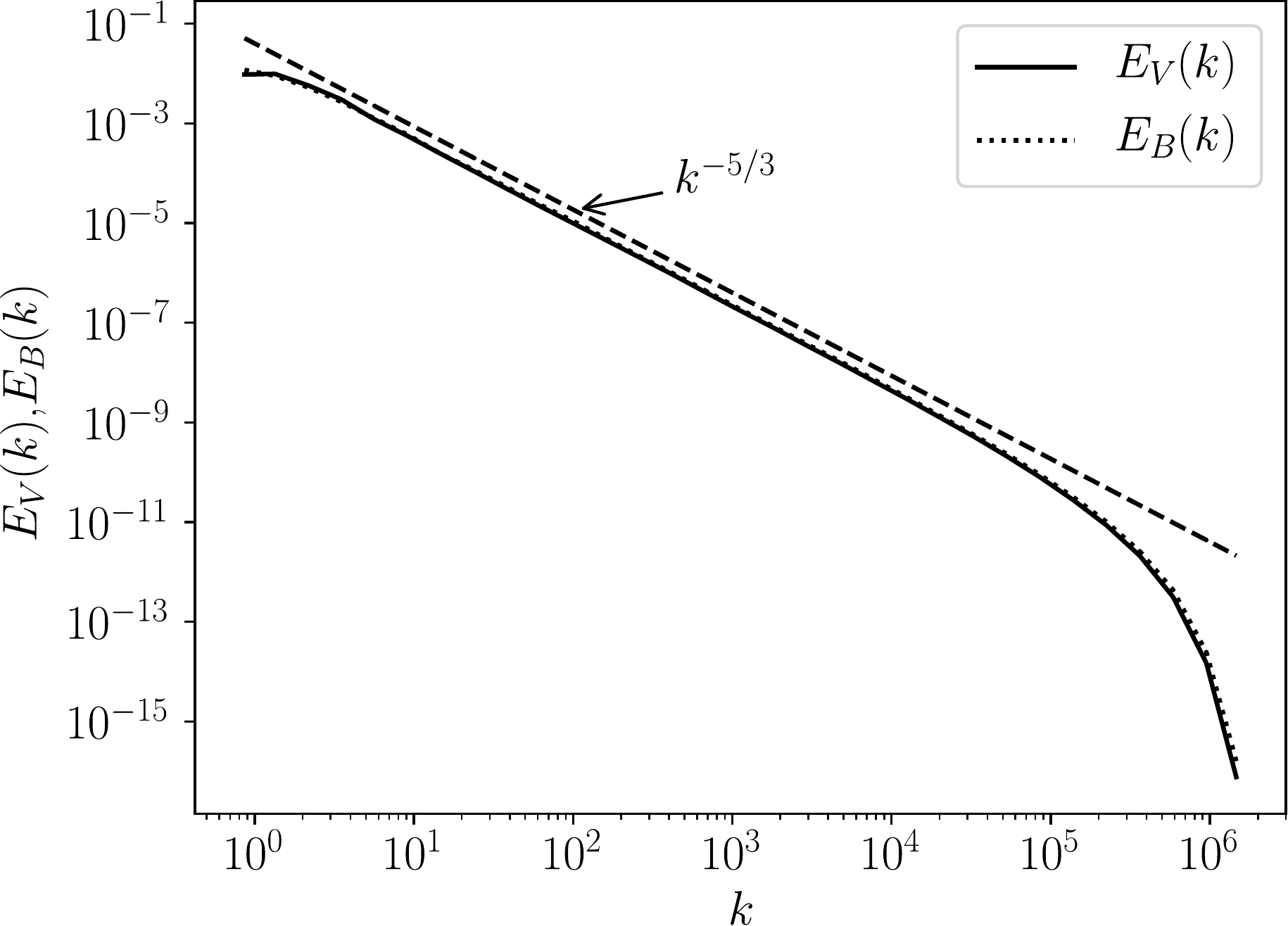}
\par\end{centering}
\caption{\label{fig:ref}Reference case with $Pr_{m}=1$, $\nu=10^{-9}$, $N=60$
and $h_{f}=10^{-3}$ kinetic (solid line) and magnetic (dotted line)
energy spectra, which follow perfectly the Kolmogorov's $k^{-5/3}$
spectrum. The case shown here is from a run up to $t=500$ and the
result is averaged over the polyhedra nodes and also from $t=460$
to $t=500$. In fact even an instanteneous spectrum is not so different
when averaged over the polyhedra nodes as shown in Ref. \citealp{gurcan:17}
for Navier-Stokes.}
\end{figure}

\paragraph{The Model-}

The nested polyhedra model of incompressible MHD equations can be
written as:
\begin{align}
\partial_{t}u_{n}^{i}+i\overline{M}_{n}^{ij\kappa}\sum_{\left\{ n',n''\right\} =\mathbf{p}_{n}}\left(u_{n'}^{\kappa*}u_{n''}^{j*}-b_{n'}^{\kappa*}b_{n''}^{j*}\right) & =-\nu k_{n}^{2}u_{n}^{i}\nonumber \\
\partial_{t}b_{n}^{i}+i\delta M_{n}^{ij\kappa}\sum_{\left\{ n',n''\right\} =\mathbf{p}_{n}}\left(u_{n'}^{\kappa*}b_{n''}^{j*}-b_{n'}^{\kappa*}u_{n''}^{j*}\right) & =-\eta k_{n}^{2}b_{n}^{i}\label{eq:model}
\end{align}
where $\overline{M}_{n}^{ij\kappa}=\left(M_{n}^{ij\kappa}+M_{n}^{i\kappa j}\right)$,
$\delta M_{n}^{ij\kappa}=\left(M_{n}^{ij\kappa}-M_{n}^{i\kappa j}\right)$
and
\[
M_{n}^{ij\kappa}\left(k\right)=k_{n}^{\kappa}\left[\delta_{ij}-\frac{k_{n}^{i}k_{n}^{j}}{k_{n}^{2}}\right]\;\text{.}
\]
Here the Einstein summation convention is used over repeated indices
and the sums are computed over the set of pairs $\mathbf{p}_{n}$
that form a triad with the node $n$, which is determined by the geometry
of nested polyhedra representation -independent from the equations-
as described in detail in Ref. \citealp{gurcan:17}. Note that if
the node belongs to the $m$th polyhedron in the nested hierarchy,
it can form triads with pairs of nodes from neighboring polyhedra
$m-2$, $m-1$, $m+1$ and $m+2$. Thus, the requirement of exact
triads and the choice of the nodes on the vertices of nested polyhedra
makes the interactions ``local'', with a constant about $62\%$
(i.e. $1/\varphi$ where $\varphi=\left(1+\sqrt{5}\right)/2$ is the
golden ratio) for the ratio between the smallest to largest wavenumber
of the interacting triad. Note that in this model this ratio is not
a seperate choice but imposed by the choice of the nested polyhedra
geometry. The notation in (\ref{eq:model}) is such that $n$ corresponds
to the node number. The node numbers $0$ to $5$ belong to the first
icosahedron (i.e. $m=0$), while $6$ to $15$ correspond to the first
dodecahedron (i.e. $m=1$) and so on. This is because we only consider
half of each polyhedron since $k_{n}^{i}\rightarrow-k_{n}^{i}$ gives
$\left\{ u_{n}^{i},b_{n}^{i}\right\} \rightarrow\left\{ u_{n}^{i*},b_{n}^{i*}\right\} $
because of the condition that the fields $\boldsymbol{u}\left(\boldsymbol{x},t\right)$
and $\boldsymbol{b}\left(\boldsymbol{x},t\right)$ should be real.
This means that in order to solve for $N$ polyhedra (i.e. ``shells''),
$8N$ nodes have to be considered. Here $n$ is the flattened node
index number, which can be defined in terms of the polyhedron index
number $m$ and the node number $\ell$ within the polyhedron in consideration.
While (\ref{eq:model}) written in such a way that the nonlinear terms
are complex conjugates, in practice the interaction defines whether
or not to complex conjugate each term. If the interaction pair in
table X has a bar it means that the corresponding term in (\ref{eq:model})
is conjugated once more, which means it goes back to the unconjugated
field. 
\begin{table*}
\begin{tabular}{|c|c|c|c|}
\hline 
$\ell^{m}$ & $\mathbf{p}_{\ell,m}=\left\{ \ell^{m-2},\ell^{m-1}\right\} $ & $\mathbf{p}_{\ell,m}=\left\{ \ell^{m-1},\ell^{m+1}\right\} $ & $\mathbf{p}_{\ell,m}=\left\{ \ell^{m+1},\ell^{m+2}\right\} $\tabularnewline
\hline 
0 & $\left\{ \left(\overline{4},\overline{0}\right),\left(\overline{5},\overline{1}\right),\left(\overline{1},\overline{2}\right),\left(\overline{2},\overline{3}\right),\left(\overline{3},\overline{4}\right)\right\} $ & $\left\{ \left(5,\overline{0}\right),\left(6,\overline{1}\right),\left(7,\overline{2}\right),\left(8,\overline{3}\right),\left(9,\overline{4}\right)\right\} $ & $\left\{ \left(\overline{5},\overline{4}\right),\left(\overline{6},\overline{5}\right),\left(\overline{7},\overline{1}\right),\left(\overline{8},\overline{2}\right),\left(\overline{9},\overline{3}\right)\right\} $\tabularnewline
\hline 
1 & $\left\{ \left(3,\overline{0}\right),\left(4,\overline{4}\right),\left(\overline{5},\overline{5}\right),\left(\overline{0},7\right),\left(\overline{2},\overline{9}\right)\right\} $ & $\left\{ \left(1,\overline{0}\right),\left(3,\overline{4}\right),\left(\overline{8},\overline{5}\right),\left(\overline{2},7\right),\left(\overline{6},\overline{9}\right)\right\} $ & $\left\{ \left(\overline{1},3\right),\left(\overline{3},4\right),\left(8,\overline{5}\right),\left(2,\overline{0}\right),\left(6,\overline{2}\right)\right\} $\tabularnewline
\hline 
2 & $\left\{ \left(5,\overline{0}\right),\left(4,\overline{1}\right),\left(\overline{3},\overline{5}\right),\left(\overline{1},\overline{6}\right),\left(\overline{0},8\right)\right\} $ & $\left\{ \left(4,\overline{0}\right),\left(2,\overline{1}\right),\left(\overline{7},\overline{5}\right),\left(\overline{9},\overline{6}\right),\left(\overline{3},8\right)\right\} $ & $\left\{ \left(\overline{4},5\right),\left(\overline{2},4\right),\left(7,\overline{3}\right),\left(9,\overline{1}\right),\left(3,\overline{0}\right)\right\} $\tabularnewline
\hline 
3 & $\left\{ \left(1,\overline{1}\right),\left(5,\overline{2}\right),\left(\overline{4},\overline{6}\right),\left(\overline{2},\overline{7}\right),\left(\overline{0},9\right)\right\} $ & $\left\{ \left(0,\overline{1}\right),\left(3,\overline{2}\right),\left(\overline{8},\overline{6}\right),\left(\overline{5},\overline{7}\right),\left(\overline{4},9\right)\right\} $ & $\left\{ \left(\overline{0},1\right),\left(\overline{3},5\right),\left(8,\overline{4}\right),\left(5,\overline{2}\right),\left(4,\overline{0}\right)\right\} $\tabularnewline
\hline 
4 & $\left\{ \left(\overline{0},5\right),\left(1,\overline{3}\right),\left(2,\overline{2}\right),\left(\overline{3},\overline{8}\right),\left(\overline{5},\overline{7}\right)\right\} $ & $\left\{ \left(\overline{0},5\right),\left(4,\overline{3}\right),\left(1,\overline{2}\right),\left(\overline{6},\overline{8}\right),\left(\overline{9},\overline{7}\right)\right\} $ & $\left\{ \left(0,\overline{0}\right),\left(\overline{4},1\right),\left(\overline{1},2\right),\left(6,\overline{3}\right),\left(9,\overline{5}\right)\right\} $\tabularnewline
\hline 
5 & $\left\{ \left(\overline{0},6\right),\left(\overline{1},\overline{8}\right),\left(2,\overline{4}\right),\left(3,\overline{3}\right),\left(\overline{4},\overline{9}\right)\right\} $ & $\left\{ \left(\overline{1},6\right),\left(\overline{5},\overline{8}\right),\left(0,\overline{4}\right),\left(2,\overline{3}\right),\left(\overline{7},\overline{9}\right)\right\} $ & $\left\{ \left(1,\overline{0}\right),\left(5,\overline{1}\right),\left(\overline{0},2\right),\left(\overline{2},3\right),\left(7,\overline{4}\right)\right\} $\tabularnewline
\hline 
\end{tabular}

\caption{\label{tab:t1}$n=8m+\ell^{m}$ is interacting with $\mathbf{p}_{n}=\left\{ n',n''\right\} =\left\{ 8m-16+\ell^{m-2},8m-10+\ell^{m-1}\right\} $,
$\left\{ 8m-10+\ell^{m-1},8m+6+\ell^{m+1}\right\} $ and $\left\{ 8m+6+\ell^{m+1},8m+16+\ell^{m+2}\right\} $
for an even $m$ (i.e. an icosahedron node) where $\ell^{m}$, $\ell^{m\pm1}$
and $\ell^{m\pm2}$ are to be taken from the values given above, where
if the integer value $n'$ has a bar over it we replace $\left\{ u_{n'}^{i*},b_{n'}^{i*}\right\} \rightarrow\left\{ u_{n'}^{i},b_{n'}^{i}\right\} $
in the interaction term in (\ref{eq:model}).}
\end{table*}
 A hybrid python/fortran numerical implementation of the model using
numpy\citep{numpy} and f2py\citep{f2py} can be found at {[}\url{http://github.com/gurcani/npm_mhd}{]}.

The model has some interesting features, such as no requirement for
the existence of a dissipative range within the simulation domain
simply by choosing the correct dissipation value. It also shows no
sign of intermittency in the sense that it follows the $S_{p}\left(k_{n}\right)\sim k_{n}^{-p/3}$
scaling in the inertial range, where $S_{p}\left(k_{n}\right)=\left\langle \frac{1}{N_{\ell}}\sum_{\ell}\left(\sum_{i}\left|u_{n\ell}^{i}\right|^{2}\right)^{p/2}\right\rangle $
and $\left\langle \cdot\right\rangle $ denotes time average.
\begin{figure}
\begin{centering}
\includegraphics[width=1\columnwidth]{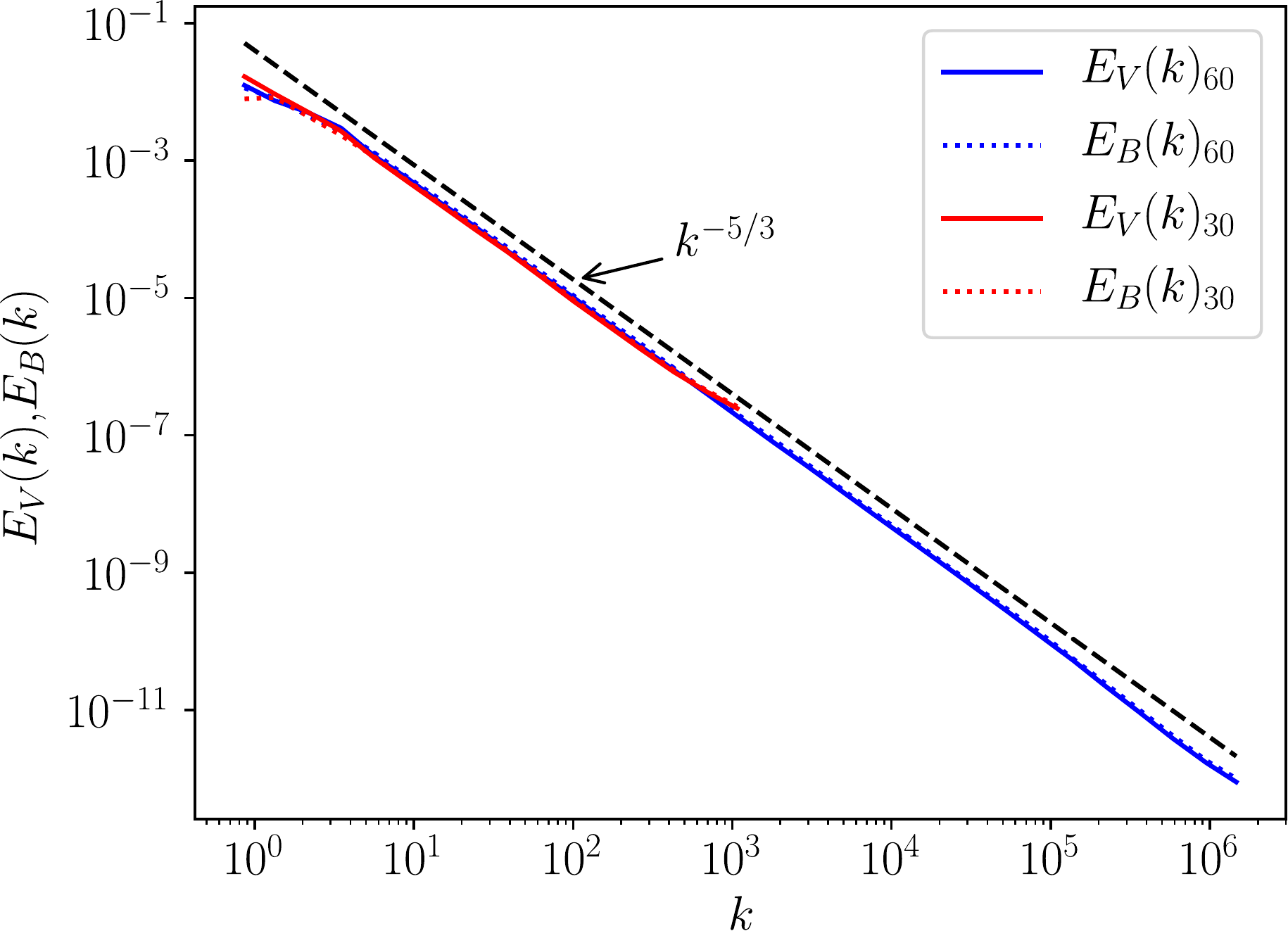}
\par\end{centering}
\caption{\label{fig:30-60}The high resolution case with $Pr_{m}=1.0$, $\nu=10^{-10}$,
$N=60$ and $h_{f}=10^{-3}$ kinetic (solid blue line) and magnetic
(dotted blue line) energies, together with the low resolution case
with $N=30$ and $\nu=10^{-6}$ kinetic (solid red line) and magnetic
(dotted red line) energies. The results are averaged over the nodes
and from $t=460$ to $t=500$.}
\end{figure}

\paragraph{Forcing-}

The model is implemented using an adaptive time step solver. Random
forcing is implemented using a fixed time step $h_{f}\sim10^{-3}$
which is larger than the maximum step size for the adaptive time stepping
algorithm. In practice the forcing is applied \emph{only on the velocity
field}, for each node $n$ of the polyhedra (i.e. shells) $m=4$ and
$m=5$ as:
\begin{figure}
\begin{centering}
\includegraphics[width=1\columnwidth]{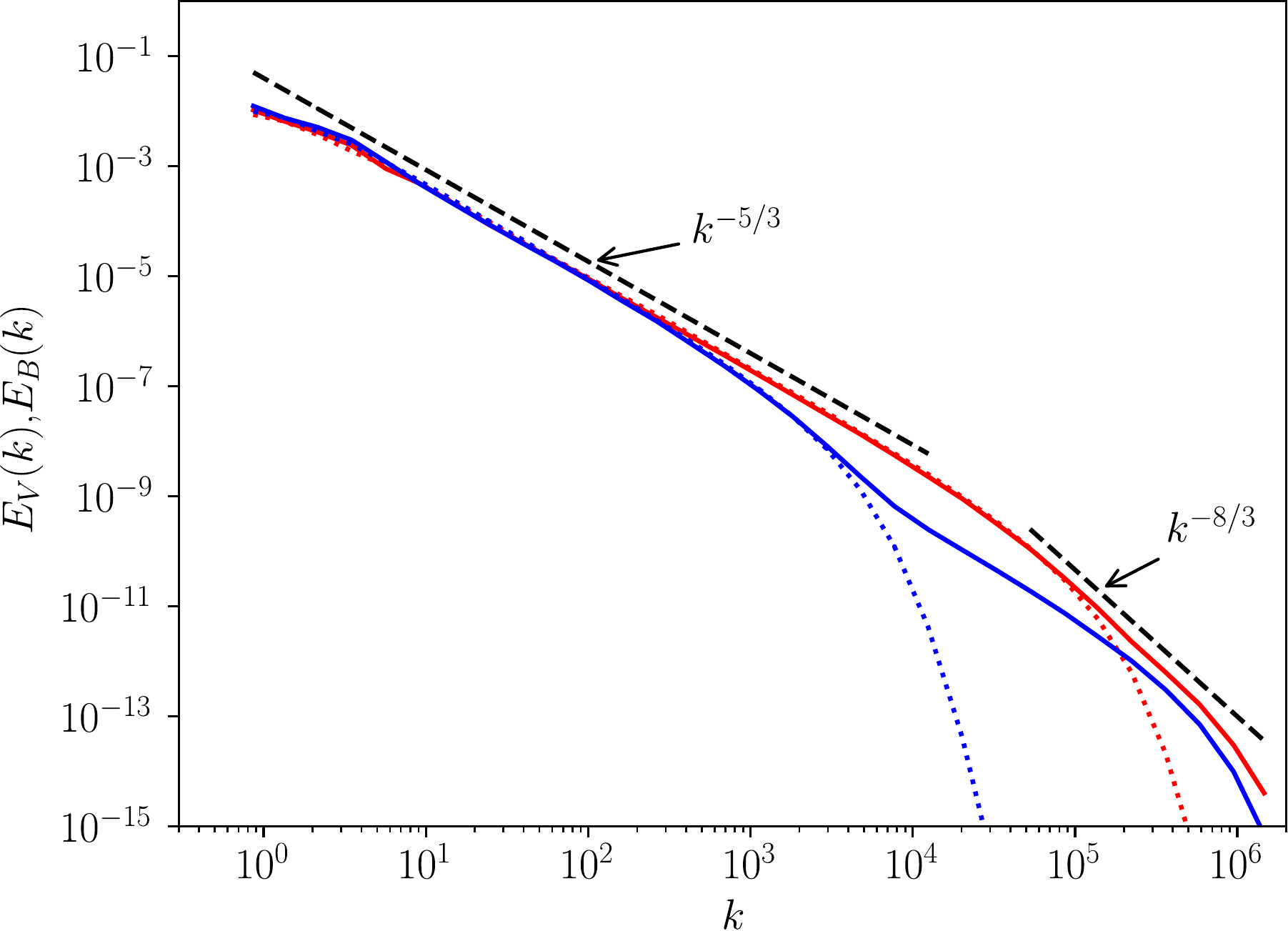}
\par\end{centering}
\caption{\label{fig:prm_small}The case with $Pr_{m}=10^{-2}$ (red) and $Pr_{m}=10^{-4}$
(blue) with $\nu=10^{-10}$, $N=60$ and $h_{f}=10^{-3}$ kinetic
(solid line) and magnetic (dotted line) energy spectra. Note that
the kinetic energy spectrum for $Pr_{m}=10^{-2}$ seems to follow
a $k^{-8/3}$ power law in the high-$k$ range where the magnetic
energy spectrum becomes dissipative. However when the Prandtl number
is decreased further this is shown to be non-universal feature. The
result is averaged over the nodes and from $t=80$ to $t=100$.}
\end{figure}
\begin{equation}
F_{n}^{i}=\left(\delta_{ij}-\frac{k_{n}^{i}k_{n}^{j}}{k_{n}^{2}}\right)\xi_{j}\label{eq:forcing}
\end{equation}
where $\xi_{j}$ is a vector random variable. The expression (\ref{eq:forcing})
guarantees that $k_{n}^{i}F_{n}^{i}=0$ and that no helicity is injected.
Indeed a preliminary attempt with $\xi_{j}=\left(1+i\right)\times10^{-2}$,
a constant, which is a standard choice in shell models, lead to the
development of large imbalances between $z^{+}=u+b$ and $z^{-}=u-b$
asymptotically. Partly expected from the fact that such a forcing
leads to strong correlation between $u$ and $b$, which modifies
the spectrum strongly\citep{grappin:83}. Question of the relation
between alignment and forcing and the relevance to real world MHD
turbulence\citep{boldyrev:06} is an important one. However, the simplest
possible mathematical approach is to choose a forcing that eliminate
velocity-magnetic field correlation\citep{mckay:17}. This urged us
to implement the random forcing discussed above, which removed the
accumulation of imbalance. In a sense the imbalance should have been
expected, since a constant forcing would lead to an accumulation of
the alignment (or anti-alignment) between $u$ and $b$.

\paragraph{Results-}

Three dimensional incompressible MHD spectra can be computed with
little difficulty up to $N=60$, where $N$ is the total number of
polyhedra in the nested polyhedra model. Starting from $k_{0}=1.0$,
one gets $k_{max}=k_{0}\varphi^{N/2}$. This means that a three dimensional
wavenumber spectrum covering a range of more than $6$ decades can
easily be simulated with such a model. This is particularly useful
if a clear identification of two or more different power laws are
desired, such as the case with large or small magnetic Prandtl numbers
$Pr_{m}\equiv\nu/\eta$.

The reference case corresponding to parameters $Pr_{m}=1$, $\nu=10^{-9}$,
$N=60$ and $h_{f}=10^{-3}$ is shown in figure \ref{fig:ref}. Indeed
this case is rather similar to a regularly discretized numerical simulation
with the same parameters, except such a run with regular discretization
would be hideously costly. One interesting aspect of the nested polyhedra
models is that the dissipative range can be eliminated as shown in
figure \ref{fig:30-60}, in this case by taking $\nu=10^{-10}$. Note
that a smaller $\nu$ with the same $N$ would lead to an increasing
spectrum around the maximum $k$. This particular feature of the nested
polyhedra model has the advantage that it does not need a subgrid
model (such as large eddy simulation or LES) to push the dissipation
range outside the simulation domain. Choosing the right value of dissipation
is sufficient. A lower resolution case with $N=30$ is also shown
in figure \ref{fig:30-60}. In fact even the case $N=30$ is sufficiently
resolved when the dissipative range is eliminated by the choice of
$\nu$. This is helpful because when one needs very good statistics
such is the case for instance, when computing structure functions
for intermittency corrections (typically runs up to $t=25000$ may
be needed) one can use lower resolution without loosing any important
features of the solution.
\begin{figure}
\begin{centering}
\includegraphics[width=1\columnwidth]{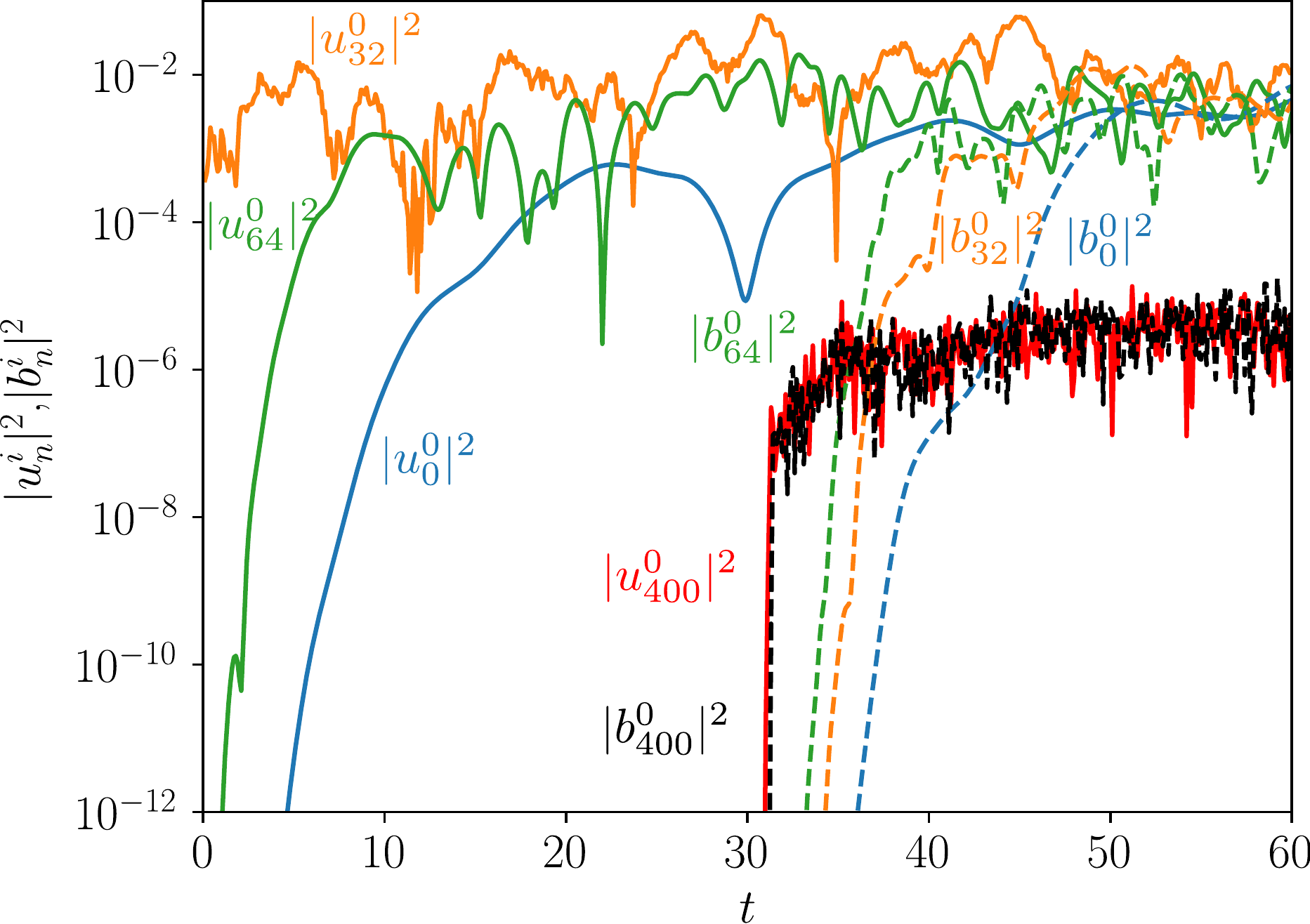}
\par\end{centering}
\caption{\label{fig:dynamo}Different modes of the system as a function of
time showing that the dynamo effect kicks in some time after the large
scales are saturated. Here we can see the effects of random forcing
on $\left|u_{32}^{i}\right|^{2}$, which then couples to other nodes.}
\end{figure}
\begin{figure}
\begin{centering}
\includegraphics[width=1\columnwidth]{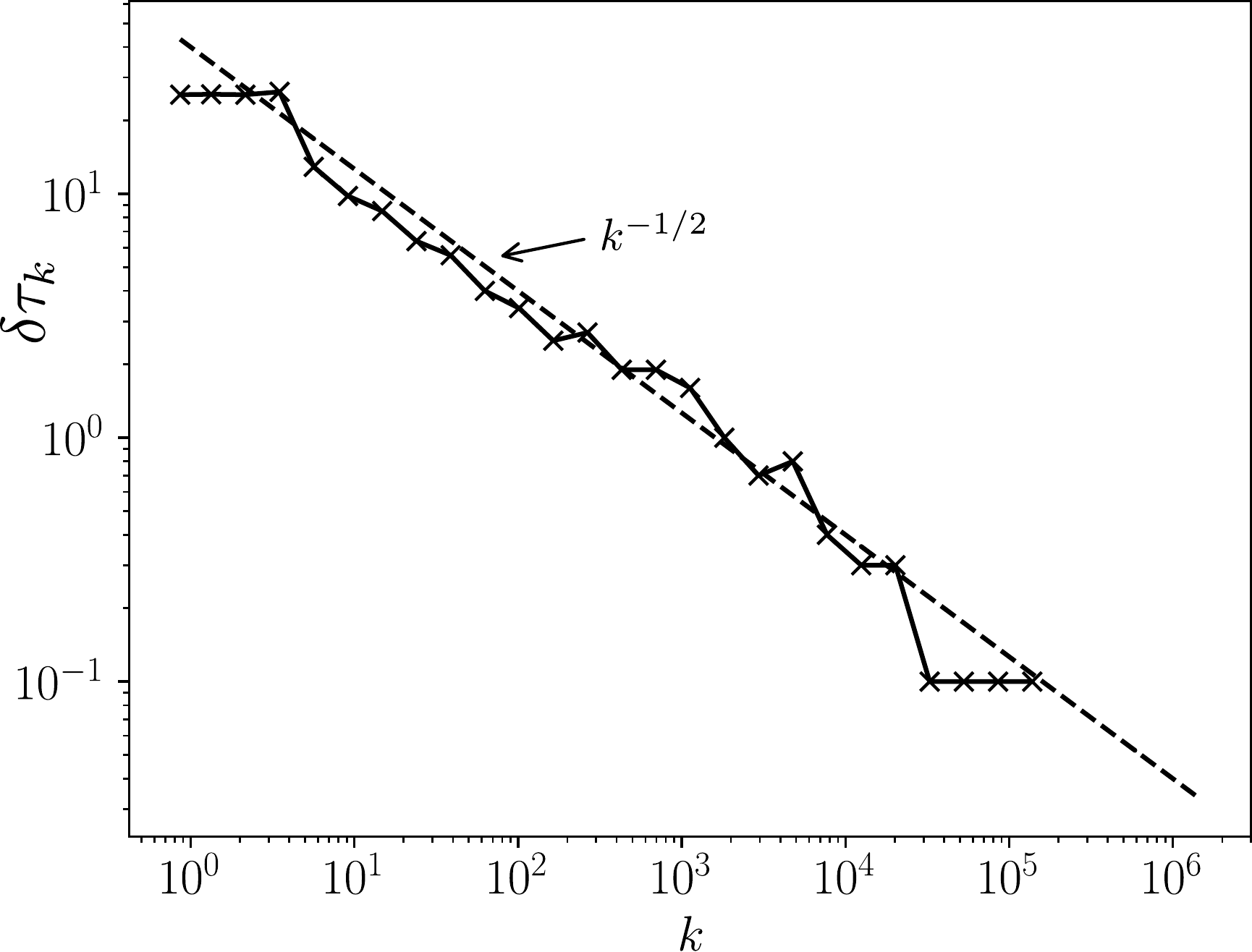}
\par\end{centering}
\caption{\label{fig:dynamo2} Equipartition time $\delta\tau_{k}=\tau_{k}-\tau_{0}$
with respect to the equipartition time of the smalles scales (i.e.
$\tau_{0}$) seems to roughly follow a $\delta\tau_{k}\propto k^{-1/2}$
scaling.}
\end{figure}
\begin{table*}
\begin{tabular}{|c|c|c|c|}
\hline 
$\ell^{m}$ & $\mathbf{p}_{\ell,m}=\left\{ \ell^{m-2},\ell^{m-1}\right\} $ & $\mathbf{p}_{\ell,m}=\left\{ \ell^{m-1},\ell^{m+1}\right\} $ & $\mathbf{p}_{\ell,m}=\left\{ \ell^{m+1},\ell^{m+2}\right\} $\tabularnewline
\hline 
0 & $\left\{ \left(\overline{5},\overline{0}\right),\left(\overline{1},\overline{1}\right),\left(\overline{4},\overline{2}\right)\right\} $ & $\left\{ \left(4,\overline{0}\right),\left(\overline{3},\overline{1}\right),\left(\overline{5},\overline{2}\right)\right\} $ & $\left\{ \left(\overline{4},\overline{5}\right),\left(3,\overline{1}\right),\left(5,\overline{4}\right)\right\} $\tabularnewline
\hline 
1 & $\left\{ \left(\overline{6},\overline{0}\right),\left(\overline{2},\overline{2}\right),\left(\overline{0},\overline{3}\right)\right\} $ & $\left\{ \left(5,\overline{0}\right),\left(\overline{4},\overline{2}\right),\left(\overline{1},\overline{3}\right)\right\} $ & $\left\{ \left(\overline{5},\overline{6}\right),\left(4,\overline{2}\right),\left(1,\overline{0}\right)\right\} $\tabularnewline
\hline 
2 & $\left\{ \left(\overline{7},\overline{0}\right),\left(\overline{3},\overline{3}\right),\left(\overline{1},\overline{4}\right)\right\} $ & $\left\{ \left(1,\overline{0}\right),\left(\overline{5},\overline{3}\right),\left(\overline{2},\overline{4}\right)\right\} $ & $\left\{ \left(\overline{1},\overline{7}\right),\left(5,\overline{3}\right),\left(2,\overline{1}\right)\right\} $\tabularnewline
\hline 
3 & $\left\{ \left(\overline{8},\overline{0}\right),\left(\overline{4},\overline{4}\right),\left(\overline{2},\overline{5}\right)\right\} $ & $\left\{ \left(2,\overline{0}\right),\left(\overline{1},\overline{4}\right),\left(\overline{3},\overline{5}\right)\right\} $ & $\left\{ \left(\overline{2},\overline{8}\right),\left(1,\overline{4}\right),\left(3,\overline{2}\right)\right\} $\tabularnewline
\hline 
4 & $\left\{ \left(\overline{9},\overline{0}\right),\left(\overline{3},\overline{1}\right),\left(\overline{0},\overline{5}\right)\right\} $ & $\left\{ \left(3,\overline{0}\right),\left(\overline{4},\overline{1}\right),\left(\overline{2},\overline{5}\right)\right\} $ & $\left\{ \left(\overline{3},\overline{9}\right),\left(4,\overline{3}\right),\left(2,\overline{0}\right)\right\} $\tabularnewline
\hline 
5 & $\left\{ \left(8,\overline{1}\right),\left(7,\overline{2}\right),\left(\overline{0},4\right)\right\} $ & $\left\{ \left(5,\overline{1}\right),\left(3,\overline{2}\right),\left(\overline{0},4\right)\right\} $ & $\left\{ \left(\overline{5},8\right),\left(\overline{3},7\right),\left(0,\overline{0}\right)\right\} $\tabularnewline
\hline 
6 & $\left\{ \left(9,\overline{2}\right),\left(8,\overline{3}\right),\left(1,5\right)\right\} $ & $\left\{ \left(1,\overline{2}\right),\left(4,\overline{3}\right),\left(\overline{0},5\right)\right\} $ & $\left\{ \left(\overline{1},9\right),\left(\overline{4},8\right),\left(0,\overline{1}\right)\right\} $\tabularnewline
\hline 
7 & $\left\{ \left(\overline{2},1\right),\left(5,\overline{3}\right),\left(9,\overline{4}\right)\right\} $ & $\left\{ \left(\overline{0},1\right),\left(2,\overline{3}\right),\left(5,\overline{4}\right)\right\} $ & $\left\{ \left(0,\overline{2}\right),\left(\overline{2},5\right),\left(\overline{5},9\right)\right\} $\tabularnewline
\hline 
8 & $\left\{ \left(\overline{3},2\right),\left(6,\overline{4}\right),\left(5,\overline{5}\right)\right\} $ & $\left\{ \left(\overline{0},2\right),\left(3,\overline{4}\right),\left(1,\overline{5}\right)\right\} $ & $\left\{ \left(0,\overline{3}\right),\left(\overline{3},6\right),\left(\overline{1},5\right)\right\} $\tabularnewline
\hline 
9 & $\left\{ \left(6,\overline{1}\right),\left(\overline{4},3\right),\left(7,\overline{5}\right)\right\} $ & $\left\{ \left(2,\overline{1}\right),\left(\overline{0},3\right),\left(4,\overline{5}\right)\right\} $ & $\left\{ \left(\overline{2},6\right),\left(0,\overline{4}\right),\left(\overline{4},7\right)\right\} $\tabularnewline
\hline 
\end{tabular}

\caption{\label{tab:t2}$n=8m+\ell^{m}+2$ is interacting with $\mathbf{p}_{n}=\left\{ n',n''\right\} =\left\{ 8m-14+\ell^{m-2},8m-4+\ell^{m-1}\right\} $,
$\left\{ 8m-4+\ell^{m-1},8m+12+\ell^{m+1}\right\} $ and $\left\{ 8m+12+\ell^{m+1},8m+18+\ell^{m+2}\right\} $
for an odd $m$ (i.e. a dodacohedron node) where $\ell^{m}$, $\ell^{m\pm1}$
and $\ell^{m\pm2}$ are to be picked from the values given above.
As in table \ref{tab:t1}, if the integer value $n'$ has a bar over
it we replace $\left\{ u_{n'}^{i*},b_{n'}^{i*}\right\} \rightarrow\left\{ u_{n'}^{i},b_{n'}^{i}\right\} $
in the interaction term in (\ref{eq:model}).}
\end{table*}

We have also considered different values of the magnetic Prandtl number
$Pr_{m}$. The case $Pr_{m}=10^{-2}$ is shown in blue in figure \ref{fig:prm_small},
representing the small magnetic Prandtl number behavior. We can see
that while there appears to be a secondary range where the magnetic
energy is dissipated and the kinetic energy seemingly displays a $k^{-8/3}$
power law scaling. However when the magnetic Prandtl number is decreased
further to $Pr_{m}=10^{-4}$, this behavior is lost and one recovers
a $k^{-5/3}$ scaling also in this range as shown in figure \ref{fig:prm_small}.
Note that, the model slows down when treating large or small Prandtl
number cases, due to explicit treatment of linear terms. It is possible
to alleviate this by using an implicit scheme or other more advance
techniques such as exponential time integration schemes. Therefore
the case with $Pr_{m}=10^{-4}$ was integrated only up to $t=100$
(saturated, but not very good statistics).

\paragraph{Dynamo-}

The simulations that are presented above, are all driven with a large
scale random forcing of the velocity field. The resulting spectra
however, present and almost perfect equipartition of kinetic and magnetic
energies. When one studies how these final steady state spectra are
established, one observes that it happens in stages. First, as the
large scale kinetic energy reaches roughly its final maximum values
a front in $k$-space of the kinetic energy appears, and fills the
whole spectral domain. As this front reaches high-$k$ end of the
inertial range (roughly about $t\approx30$ for the reference case
above), equipartition between kinetic and magnetic energies gets established
at high-$k$. Then another front (this time of the magnetic energy
density) fills up the $k$-range moving towards smaller $k$. We can
define the time it takes for the establishment of the equipartition
$\tau_{k}$ which is a function of the wave-number $k$, which in
general is a function of the initial conditions. The case of the very
small seed initial conditions are shown in figure \ref{fig:dynamo2}.

\paragraph{Conclusion-}

We show that a nested polyhedra model, obtained from ``decimating''
the wave-number space using self-similarly scaled nested, alternating
icosahedra and dodecahedra, such that the wave-vectors that corresponds
to two nodes of the system can combine to give a third one that also
falls on a resolved node, can be used to model the MHD system of equations
with no external magnetic field. In this model, the interactions are
``local'' in $k$-space (i.e. the ratio $k_{n-2}/k_{n}$ of the
smallest to the largest wavenumbers of the interacting triad is about
$62\%$).

Considering isotropic MHD turbulence with no background magnetic field
or rotation, and random large scale forcing on the velocity component,
we find that the model can display a clear Kolmogorov power law scaling
of the form $k^{-5/3}$ over $6$ decades in wave-number space with
very good statistics, which allows considering large or small magnetic
Prandtl number cases. Moreover, with a careful choice of the high-$k$
dissipation, the apparent inertial range can extend all the way up
to the end of the resolved range in $k$-space due to perfect self-similarity.
Finally, since the random forcing was applied only on velocity, the
magnetic energy spectrum gets established via the dynamo effect that
starts from the small scales. It was observed that the time scale
$\delta\tau_{k}=\tau_{k}-\tau_{0}$ for the equipartition, offset
with the time of equipartition of the smallest scales shows a $k^{-1/2}$
scaling.
\begin{acknowledgments}
The author would like to thank W.-C. Müller, P. Morel, P. H. Diamond,
R. Grappin and attendants of the \emph{Festival de Théorie, Aix en
Provence} in 2017.
\end{acknowledgments}

\end{document}